\documentclass[aip,jcp,preprint,floatfix]{revtex4-1}
\usepackage[english]{babel}
\usepackage{amsmath,amsthm}
\usepackage{graphicx}
\usepackage{subfigure}
\usepackage{amsfonts}

\theoremstyle{definition}

\theoremstyle{remark}

\numberwithin{equation}{section}
\begin{document}
\title{The Gaussian Diffusion Approximation for Complex Fluids is Generally Invalid}
\author{George D. J. Phillies}
\email{phillies@wpi.edu}
\affiliation{Department of Physics, Worcester Polytechnic Institute, Worcester, MA, 01609}
\date{\today}%
\begin{abstract}
Simulations are made of a probe particle diffusing through a complex fluid.  Probe particle motions are described by the Mori-Zwanzig equation and Mori's orthogonal hierarchy of random forces scheme, subject to  the approximation that the fluid creates a rapidly-fluctuating random force corresponding to solvent motions and a slowly fluctuating random force corresponding to solute (e.~g., matrix polymer) motions. The Gaussian diffusion approximation is seriously incorrect in this physically-plausible model system. $P(x,t)$ has exponential wings. $g^{(1s)}(q,t)$ can differ from $\exp(-q^{2} \langle x^{2}\rangle/2$ by up to orders of magnitude. Experimental interpretations that rely on the Gaussian approximation, such as the Stejskal-Tanner equation for pulsed-field-gradient NMR or particle tracking, can not be assumed to be reliable in complex fluids.
\end{abstract}
\maketitle
\section{Introduction}

There is a long-time\cite{hallet1974a,turner1976a} interest in studying the properties of complex fluids by observing the diffusion through them of dilute mesoscopic probes. Experimental studies have applied a multitude of different experimental techniques and disparate interpretational approaches. As discussed below, experimental techniques for studying probe diffusion have included quasielastic light scattering, fluorescence recovery after photobleaching, fluorescence correlation spectroscopy, pulsed-field-gradient nuclear magnetic resonance, inelastic neutron scattering, diffusing wave spectroscopy, and optical particle tracking. These experiments are complemented by computational studies using on-lattice Monte Carlo and off-lattice molecular dynamics simulations. Many interpretations of probe diffusion measurements explicitly or implicitly invoke the Gaussian diffusion approximation. In some experimental studies, the Gaussian approximation is taken as a given.  In others, the Gaussian diffusion approximation is taken to be invalid, or is ignored as not being relevant to the data analysis.

The Gaussian diffusion approximation for one-dimensional diffusion is\cite{berne1976a,phillies2000a}
\begin{equation}
    P(\Delta x, t) = \exp(- (\Delta x)^{2}/(2\langle (\Delta x(t))^{2}\rangle )),
    \label{eq:gaussianapprox}
\end{equation}
where $P(\Delta x, t)$ is the probability of observing a displacement $\Delta x$ of a diffusing particle during a time interval $t$, and where $\langle (\Delta x(t))^{2}\rangle$ is the mean-square displacement of the diffusing particle during time $t$.  The mean-square displacement is in turn related to the diffusion constant $D$ via
\begin{equation}
    \langle (\Delta x(t))^{2}\rangle = 2 D t.
    \label{eq:meansquareD}
\end{equation}

The relationships between eqs \ref{eq:gaussianapprox} and \ref{eq:meansquareD} and probe diffusion are generally taken to arise from the Langevin equation, the central limit theorem, and Doob's theorem\cite{doob1942a}, as discussed, e.g., in Berne and Pecora\cite{berne1976a}.  The Langevin equation is a heuristic approximation for the equation of motion of a diffusing particle. Written in one dimension, the Langevin equation provides
\begin{equation}
    m \frac{d^{2} x}{dt^{2}} = - f_{o} \frac{dx}{dt}  +{\cal F}(t),
    \label{eq:langevin}
\end{equation}
Here $x$ is the time-dependent position of the diffusing particle, $m$ and $f_{o}$ are the probe's mass and drag coefficient, and ${\cal F}(t)$ is the random (thermal) force on the particle. $f_{o}$ and ${\cal F}(t)$ are not independent; they are interlinked by the requirements that the probe's mean-square velocity satisfies the equipartition theorem and has no long-term secular drift. ${\cal F}(t)$ is generally taken to be a Gaussian random Markoff process. Standard solutions of the Langevin equation\cite{phillies2000a} show that nonoverlapping probe displacements are identically distributed and very nearly independent.

The central limit theorem describes a random variable, a variable whose values are generated by adding together a large number of identically distributed, uncorrelated, random steps having average value zero. Doob's theorem\cite{doob1942a} describes a random \emph{process}, a time-dependent variable  whose changes in value from each time to the next are generated by adding together a large number of identically distributed, uncorrelated, random steps having average value zero. The successive displacements $\Delta x(t)$ of a diffusing probe, whose motions obey the Langevin equation, satisfy the requirements of the central limit theorem and equally satisfy the requirements of Doob's theorem. The central limit theorem guarantees that $P(\Delta x, t)$ follows eq \ref{eq:gaussianapprox}, namely $P(\Delta x, t)$ must be a Gaussian in $\Delta x$.  Doob's Theorem equally guarantees that $\langle (\Delta x(t))^{2}\rangle$ must follow eq \ref{eq:meansquareD}, namely $\langle (\Delta x(t))^{2}\rangle$ must increase linearly with time.

The Langevin equation, and the Gaussian diffusion approximation that follows from it, were constructed as an approximate description of small probes diffusing in simple Newtonian fluids. More recently\cite{hallet1974a,turner1976a}, experiment has advanced to the study of objects diffusing in complex fluids, fluids such as nondilute solutions of colloids, polymers, proteins, or surfactants, not to mention intracellular media \emph{in vivo}. Physically, complex fluids are characterized by relaxations on a wide range of time scales. When measurements on diffusing probes in complex fluids extend to sufficiently short times, the figurative individual steps that add together to describe probe motion cease to be independent; over short times the steps are correlated. The rationales leading to the Langevin equation, the central limit theorem, and Doob's theorem then cease to be applicable.

Experiment confirms that the Gaussian diffusion approximation, as arising from the central limit theorem and Doob's theorem, is not generally valid in complex fluids:

First, there are direct measurements of $P(\Delta x, t)$ using particle tracking. Early measurements of $P(\Delta x, t)$ by Apgar, et al.\cite{apgar2000a} and Tseng, et al.\cite{tseng2001a} clearly revealed non-Gaussian forms for $P(\Delta x, t)$. More recent studies by Wang, et al.\cite{wang2009a,wang2012a}, and Guan, et al.\cite{guan2014a}, the last being measurements on colloidal hard spheres diffusing through nondilute suspensions of larger hard spheres, not only confirm a non-gaussian distribution of $P(\Delta x, t)$ but reveal its form, namely $P(\Delta x, t)$ is nearly Gaussian for smaller $\mid \Delta x \mid$, but at larger $\mid \Delta x \mid$ decreases exponentially in $\mid \Delta x \mid$.

Second, $\langle (\Delta x(t))^{2}\rangle$ can be measured directly using particle tracking. In some systems, experiment finds
\begin{equation}
     \langle (\Delta x(t))^{2}\rangle = a t^{\alpha}
     \label{eq:nondiffusion}
\end{equation}
for $a$ a constant and $\alpha \neq 1$; the case $\alpha < 1$ is termed subdiffusion. If subdiffusion is observed, the outcome guaranteed by Doob's theorem is not being obtained. Particle motion is then mathematically certain not to be described by a Gaussian-random Markoff process, because if it were Gaussian-random Markoff, $\alpha =1$ would with mathematical certainty by obtained. Correspondingly, $P(\Delta x, t)$ for subdiffusive systems cannot be be Gaussian.

Third, \emph{in probe systems} some experimental techniques, e.g., light scattering spectroscopy or inelastic neutron scattering, measure directly the spatial Fourier transform of $P$, namely
\begin{equation}
      g^{(1s)}(q,t) = \int_{-\infty}^{\infty} d \Delta x \ P(\Delta x, t) \exp(i q \Delta x).
      \label{eq:dynamicstructure}
\end{equation}
Eq \ref{eq:dynamicstructure} does not apply in non-probe systems in which the scatterers are not dilute. For light scattering spectroscopy on probe systems, up to a possible normalizing constant not significant here, $g^{(1s)}(q,t)$ is the self part of the dynamic structure factor, with $q$ being the scattering vector. The Gaussian approximation predicts that eq \ref{eq:dynamicstructure} becomes
\begin{equation}
      g^{(1s)}(q,t) = \exp(- q^{2} \langle (\Delta x(t))^{2}\rangle/2).
      \label{eq:dynamicstructure2}
\end{equation}
Eq \ref{eq:dynamicstructure2} is sometimes interpreted as suggesting that $\langle (\Delta x(t))^{2}\rangle$ can in general be extracted from $g^{(1s)}(q,t)$. However, if eq \ref{eq:dynamicstructure2} is correct, Doob's theorem guarantees
\begin{equation}
    g^{(1s)}(q,t) = \exp(- q^{2} D t);
    \label{eq:g1sqtbrownian}
\end{equation}
in this case $\log(g^{(1s)}(q,t))$ is linear in $q^{2}$ and $t$.

QELSS studies of probe diffusion in complex fluids readily identify systems in which eq \ref{eq:g1sqtbrownian} is incorrect.  In some systems\cite{streletzky1998b}, $g^{(1s)}(q,t)$  relaxes as a stretched exponential $\exp(- \theta t^{\beta})$, $\theta$ and $\beta$ being line shape parameters, with $\beta \neq 1$.  In other systems, $g^{(1s)}(q,t)$ gains multiple relaxations on different time scales\cite{dunstan2000a,bremmell2001a,streletzky1998a}. In some systems\cite{dunstan2000a,bremmell2001a,streletzky1998a,phillies2003a}, the relaxation $\log(g^{(1s)}(q,t))$ is not linear in $q^{2}$. In each of these systems, the Gaussian diffusion approximation cannot possibly be valid, because $g^{(1s)}(q,t)$ has properties not consistent with the central limit theorem and Doob's theorem.

Based on experiment, the Langevin equation and the Gaussian diffusion approximation must not be uniformly valid for probes diffusing in complex fluids.

A variety of mathematical paths have been advanced to extend beyond the Langevin equation. Mandelbrot and Van Ness\cite{mandelbrot1968a} discuss fractional Brownian motion. Fractional Brownian motion differs from the Brownian motion generated by the Langevin equation (eq \ref{eq:langevin}) in that the simple random force ${\cal F}(t)$ is replaced with an integral average
\begin{equation}
    {\cal F}_{M}(t) = \int ds K(s) {\cal F}(t-s)
     \label{eq:FBMrandomforce}
\end{equation}
of random forces applied at different times, $K(s)$ being a memory kernel. The simple random force had a vanishingly short correlation time, so that $\langle {\cal F}(t){\cal F}(t+s)
\rangle \sim \delta(s)$, $\delta(s)$ being the Dirac delta function.  In fractional Brownian motion, $K(s)$ is non-zero over an extended range of values of $s$, so that the random increments supplied to $dx/dt$ by ${\cal F}_{M}(t)$ at different times are cross-correlated. Mandelbrot and Van Ness\cite{mandelbrot1968a}  specifically considered a power-law memory kernel.  So long as ${\cal F}_{M}(t)$ is a sum of \emph{identically distributed} Gaussian random variables, it is itself a Gaussian random variable, so the distribution of displacements $P(\Delta x,t)$ generated by fractional Brownian motion remains Gaussian.

Closely related to fractional Brownian motion are the motions described by the generalized Langevin equation
\begin{equation}
    m \frac{d^{2} x}{dt^{2}} = - \int_{-\infty}^{t} ds\ \phi(t-s) \frac{dx(s)}{dt}  +{\cal F}(t),
    \label{eq:langevingeneral}
\end{equation}
as discussed by Fox\cite{fox1977a}, with memory kernel $\phi(t-s) = k_{B} T m^{2} \langle {\cal F}(t) {\cal F}(s) \rangle$, and $k_{B}$ and $T$ being Boltzmann's constant and the absolute temperature, respectively. The random force ${\cal F}(t)$ is taken to be a non-Markoffian Gaussian random process, non-Markoffian because $\phi(t) \neq a \delta(t)$. As emphasized by Fox\cite{fox1977a}, $dx(t)/dt$ inherits from ${\cal F}(t)$ its Gaussian-random non-Markoff nature, so that $x(t)$ in turn is a Gaussian non-Markoffian process.

An alternative to fractional Brownian motion is provided by the continuous-time random walk, in which the diffusing particle takes identically-distributed Gaussian-random steps, but in which the nominal time interval associated with each step is a separately-determined identically-distributed random number. A different alternative to fractional Brownian motion is provided by simple diffusion, in which diffusion is confined to a percolation cluster at the threshold. The continuous time random walk has no characteristic time scale.  The percolation cluster has no characteristic distance scale. Saxton\cite{saxton2001a} demonstrates the effect of these non-Langevin diffusion processes on relaxation curves from fluorescence recovery after photobleaching.

Walks generated by fractional Brownian motion, the generalized Langevin equation, and the continuous-time random walk have in common the outcome that the walk is a Gaussian random process, which is completely characterized by a single two-time correlation function $\langle (\Delta x(t))^{2} \rangle$. Correspondingly, many studies\cite{schwille1999a,saxton2001a} of subdiffusive motion, whether pursued experimentally or by computer simulation, have focused on determining $\langle (\Delta x(t))^{2} \rangle$, which for Gaussian random processes whether Markoffian or not suffices to characterize the process completely. However, experimental results noted above show conclusively that $P(\Delta x, t)$ from probes in complex fluids is not in general a Gaussian in $\Delta x$, so probe diffusion in complex fluids must not correspond to any of these mathematical processes.

The objective of this paper is to demonstrate an alternative theoretical treatment of probe diffusion in complex fluids that has a sound physical basis and that generates diffusive processes that agree with experiment. There is no claim that our treatment is unique. The next section of the paper develops the theoretical and computational basis of our solution, including a discussion of various computational diagnostics that give information on aspects of $P(\Delta x, t)$. Section III of the paper presents an exemplary simulation, leading to a $P(\Delta x, t)$ that agrees with the experiments of Wang\cite{wang2012a}, Guan\cite{guan2014a}, and co-workers. Section IV discusses implications of our work for various experimental techniques that have been applied to study probe diffusion.

\section{Theoretical Background}

Our results are based on the Mori-Zwanzig equation\cite{mori1965a}, which is an exact rearrangement of the physically-exact Liouville equation for the time evolution of classical systems. The Mori-Zwanzig equation \cite{mori1965a} provides
\begin{equation}
      m \frac{d u(t)}{dt}  =   i \Omega u(t)- \int_{-\infty}^{t} ds \phi(s) u(t-s) + F^{P}(t).
       \label{eq:MoriZwanzig}
\end{equation}
Here $u(t)$ is the dynamic variable of interest, in this work the probe velocity. For our systems $\Omega$ vanishes by time reversal symmetry. $F^{P}(t)$ is the Mori-Zwanzig projected force. The Mori-Zwanzig theorem gives an exact expression for $F^{P}(t)$ in terms of the system Hamiltonian. The Mori memory kernel is
\begin{equation}
     \phi(s) = \langle F^{P}(0) F^{P}(s) \rangle / \langle (u(0))^{2} \rangle.
     \label{eq:morimemorykernel}
\end{equation}
The Mori-Zwanzig equation thus replaces the Langevin equation and gives an exact -- albeit difficult to evaluate -- formula for the memory kernel $\phi(s)$. Equations \ref{eq:langevin} and \ref{eq:MoriZwanzig} are fundamentally different. Eq \ref{eq:langevin} is often interpreted as a stochastic differential equation. Eq \ref{eq:MoriZwanzig} is a conventional differential equation: It is Newton's second law of motion, rewritten by partitioning the force on the probe particle due to the other molecules in the system between $\Omega$, $\phi(s)$, and $F^{P}(t)$. $F^{P}(t)$ is determined by the positions and motions of the other particles in the system, so it is continuous, differentiable, and integrable. Difficulties associated with integrating stochastic differential equations\cite{doob1942a} do not arise with the Mori-Zwanzig equation.

$F^{P}(t)$ is often approximated as having a correlation time short compared to the time scales of interest, so that $F^{P}(t)$ is approximated by a Markoff process, while $\phi(s)$ can be approximated as being very nearly $\sim \delta(s)$. For the systems under consideration here, these approximations would lose all the interesting physics. The central interest in observing probe diffusion in complex fluids is to extract information about relaxations of the complex fluids. To do so, probe motions must be observed on the time scales on which relaxations occur. On these time scales, $F^{P}(t)$ is not even approximately a Markoff process; it instead has prolonged correlations related to the prolonged correlations in the surrounding fluid.

Mori has supplied a useful computational approach for generating an $F^{P}(t)$ with well-defined correlations, together with a mutually consistent $\phi(t)$, namely the orthogonal hierarchy of thermal forces scheme\cite{mori1965b}. The basis of the orthogonal hierarchy is Mori's observation that the Mori-Zwanzig equation is valid for an arbitrary dynamic mechanical variable, the thermal force $F^{P}(t)$ is a dynamic mechanical variable, so therefore the time evolution of $F^{P}(t)$ can itself be calculated with a new Mori equation. The new Mori equation generates the time evolution of $F^{P}(t)$ in terms of a second Mori memory kernel and a second thermal force. Each thermal force $F^{P}(t)$ can in turn be written as being generated by a higher-order memory kernel and thermal force. The orthogonal hierarchy automatically leads to a mutual consistency between $\phi$ and $F^{P}(t)$. Here we use the orthogonal hierarchy purely as a mathematical device to generate an $F^{P}(t)$ that has the desired time correlations and a $\phi(t)$ whose time dependence is consistent with the time correlations in $F^{P}(t)$. Our device is to truncate the hierarchy at some order, and then use the highest-order equation purely as a generalized Langevin equation that yields projected forces and corresponding memory functions having the desired temporal calculations.

In the following calculations a complex fluid is modelled as supplying two independent projected forces. One is a rapidly-fluctuating solvent force corresponding to the simple hydrodynamic drag $- f_{o} v$ on the probe.  The other is a slowly-fluctuating projected force corresponding to complex fluid (e.~g., dissolved polymer matrix) motions. The presence of two distinct projected forces is critical to obtaining our results.

Our approach is mathematically closely related to the treatment of Tateishi, et al.\cite{tateishi2012a}, who considered a generalized Langevin equation containing two uncorrelated noise sources $\xi(t)$ and $\eta(t)$. The time correlation functions of $\xi(t)$ and $\eta(t)$ were a delta function and a power law. Because $\xi(t)$ and $\eta(t)$ were uncorrelated, the corresponding memory kernel was
\begin{equation}
    \phi(\tau) = \langle \xi(t)\xi(t+\tau)\rangle + \langle \eta(t) \eta(t+\tau) \rangle.
    \label{eq:tate2}
\end{equation}
$\xi(t)$ and $\eta(t)$ have different distributions, so their sum is not a sum of identically distributed random variables; the central limit theorem and Doob's theorem are therefore not applicable to their sum. Tateishi's analytic calculation of $ \langle (\Delta x (t))^{2} \rangle$, based on this model, found distinct diffusive and subdiffusive regimes. The calculations here differ from those of Tateishi in that we calculated $P(\Delta x, t)$ itself, and furthermore calculated a range of statistical characterizations and transforms of $P(\Delta x, t)$.

Simulations were run on an 448 core Nvidia Tesla C2075 processor (nominal maximum single-precision speed, 1.15 teraflops) using the Portland Group PGFortran optimizing compiler for Fortran 90. Individual simulations ran for $5 \cdot 10^{9}$ particle displacement steps. The direct outcomes of each simulation were a position trajectory $x_{i}$ and a velocity trajectory $u_{i}$, $i$ being the discrete time variable. $u_{i}^{2}$ was confirmed to have no secular drift over the course of a simulation, confirming that the system remained in thermal equilibrium.

Throughout the simulations, changes in the position were computed from the $u_{i}$ as
\begin{equation}
    x_{i} = x_{i-1} + u_{i} \Delta t.
    \label{eq:displacementvel}
\end{equation}
In final simulations, notional units were chosen so that $\Delta t =1$.

A simulation of the Langevin equation was made as a final software test. For the Langevin simulation, the Langevin equation for the velocity was used in its discrete-time form
\begin{equation}
     u_{i} = u_{i-1} - f_{o} u_{i-1} \Delta t  + X_{i} \Delta t
     \label{eq:Langevinstepped}
\end{equation}
Here $i$ labels the time steps. $X_{i}$ is a net impulse, the integral of the projected force over the time interval between moments $i-1$ and $i$. In the simulations, $X_{i}$ and $X_{j}$ for $i \neq j$ were independently generated Gaussian random variables. Because time is discretized, the $X_{i}$ mathematically cannot have a correlation time shorter than $\Delta t$.

For the complex fluid simulation, we added to the Langevin equation a second projected force, one having an extended correlation time, and its corresponding memory function. The projected force was constructed as a sum over Markoff sources $Y_{j}$, the effect of these sources being propagated forward from time $j$ to time $i$ by propagators $C_{i-j}$, namely
\begin{equation}
      F^{P}_{i} = \sum_{j=0}^{i}  Y_{j} C_{i-j}
      \label{eq:Morirandomforce}
\end{equation}
The propagators $C_{i-j}$ have a range $N$, meaning that they are only non-zero for $\mid i-j \mid < N$. During the course of a simulation $i \gg N$. Because each $Y_{j}$ contributes to a substantial number of  ${\cal F}_{i}$, the ${\cal F}_{i}$ are cross-correlated. Because $F^{P}_{i}$ is constructed as a sum of Gaussian random processes, the probability distribution of $F^{P}_{i}$ must also be a Gaussian random process, as was confirmed in the simulations.  However, $C_{i-j}$ is non-zero for $i-j \neq 0$, so the long-lived $F^{P}_{i}$ are cross-correlated; the long-lived $F^{P}_{i}$ are not a Markoff process.

From eq \ref{eq:morimemorykernel}, the Mori kernel for the second projected force may be written in terms of the propagator as
\begin{equation}
     M_{b-a} \equiv \langle {\cal F}_{a} {\cal F}_{b} \rangle  = \langle \sum_{i=1}^{a} \sum_{j=1}^{b}  Y_{i} C_{a-i}Y_{j} C_{b-j}\rangle,
      \label{eq:Morirandomforce2}
\end{equation}
with $a \gg N$ and $b \gg N$. For $b-a \geq 0$ and
\begin{equation}
    \langle Y_{i} Y_{j} \rangle = m_{1}^{2} \delta_{i-j},
    \label{eq:Ycorrelation}
\end{equation}
with $\delta_{i-j}$ being the Kronecker delta, $M_{b-a}$ simplifies to
\begin{equation}
     M_{j} = m_{1}^{2} \sum_{i=0}^{N} C_{i} C_{i+j}.
         \label{eq:Morirandomforcefinal}
\end{equation}
Our propagator was an exponential
\begin{equation}
   C_{i} = f_{1} \exp(-a i)/Q
   \label{eq:exppropagator}
\end{equation}
with $f_{1}$ being the strength of the propagator, the normalizing factor $Q$ being arranged for each propagator so that
\begin{equation}
      f_{1} = \sum_{i=0}^{N} C_{i}.
   \label{eq:exptotal}
\end{equation}
By direct calculation, for an exponential propagator the memory kernel is also an exponential, namely
\begin{equation}
     M_{j}   = \left[ f_{1}^{2} \sum_{i=0}^{N} \exp(-2 a i)/Q^{2} \right] \exp(-a j)
     \label{eq:expmemory}
\end{equation}
the quantity in brackets being a constant independent of $j$.

We also tested propagators that initially followed eq \ref{eq:exppropagator}, but at times $i > a^{-1}$ followed a power law
\begin{equation}
      C_{i} = f_{1} (i a)^{\nu} /(Q e),
      \label{eq:powerlaw}
\end{equation}
with $f_{1}$ chosen so that $C_{i}$ was continuous at the crossover point.

The discrete-time Mori equation becomes
\begin{equation}
    u_{i}  = (u_{i-1}+ X_{i} +\sum_{j=0}^{N} \left(C_{j} Y_{i-j} - M_{j} u_{i-j}\right)) (1-f_{o}).
    \label{eq:morispecific}
\end{equation}
The $u_{i}$ are driven by  two different statistical processes, one having an extended memory, so neither the Central Limit Theorem (which requires a sum of \emph{identical} processes) nor Doob's Theorem (which refers to Markoff processes) is applicable to the behavior of the $u_{i}$.

Having generated the statistical processes $u_{i}$ and $x_{i}$ for 9 billion steps (plus initial thermalization), characterizations of these processes followed. For ease of reading, the characterizations are written with time as the continuous variable $t$. For each system we calculated the displacement distribution function $P(\Delta x,t)$, the velocity-velocity correlation function
 \begin{equation}
     C_{VV}(t) = \langle u(0) u(t) \rangle,
      \label{eq:CVV}
\end{equation}
and the acceleration-acceleration correlation function
\begin{equation}
     C_{AA}(t) = \langle (u(t_{2})- u(t_{1}))(u(t_{4})- u(t_{3})) \rangle.
      \label{eq:DVDV}
\end{equation}
Here $\Delta x(t) = x(\tau+t) - x(\tau)$. The function $C_{AA}(t)$ was evaluated for $t_{1} \leq t_{2} \leq t_{3} \leq t_{4}$, with $t = t_{3} - t_{2}$, while keeping $t_{2}- t_{1}$ and $t_{4}- t_{3}$ small.  $P(\Delta x, 1)$ gives the distribution of $x_{i}-x_{i-1}$, which is the same as the distribution of the $u_{i}$. The $u_{i}$ had the expected Gaussian distributions.

The velocity-velocity correlation functions are long lived, so errors in eq \ref{eq:displacementvel} due to time being discretized were small. For the simple Langevin model, the velocity-velocity correlation function was accurately exponential, demonstrating $f_{o} \Delta t$ was not too large. The time-dependent mean-square displacement
\begin{equation}
     K_{2}(t) = \langle (\Delta x(t))^{2} \rangle
      \label{eq:K2t}
\end{equation}
was computed directly, not from $P(\Delta x,t)$. Plots of $P(\Delta x, t)$ were generated by binning values of $P(\Delta x, t)$ using $0.1 \sqrt(K_{2}(t))$ as the bin width.

Unless $P(\Delta x,t)$ is a Gaussian, characterizing $P(\Delta x,t)$ requires all even central moments $K_{2n}$ of $\Delta x$. We calculated the time-dependent $K_{4}$ and $K_{6}$ from the simple moments as
\begin{equation}
      K_{4} =  (\langle (\Delta x(t))^{4} \rangle -3(\langle (\Delta x(t))^{2} \rangle)^{2})/(\langle (\Delta x(t))^{2} \rangle)^{2}
      \label{eq:K4t}
\end{equation}
and
\begin{displaymath}
      K_{6} = (\langle (\Delta x(t))^{6} \rangle -15 \langle (\Delta x(t))^{4} \rangle\langle (\Delta x(t))^{2} \rangle
\end{displaymath}
\begin{equation}
       +30\langle (\Delta x(t))^{2} \rangle^{3})/(\langle (\Delta x(t))^{2} \rangle)^{3}.
      \label{eq:K6t}
\end{equation}
The odd central moments  $K_{1}$, $K_{3}$, and $K_{5}$ of $P(\Delta x,t)$ were confirmed by direct calculation to vanish, as expected from symmetry.

The intermediate scattering function
\begin{equation}
      g^{(1s)}(q,t) = \langle \cos( q \Delta x(t)) \rangle
      \label{eq:g1s}
\end{equation}
was determined for a wide range of $q$ and $t$. As an indication of the simulation's accuracy, the relaxation of the dynamic structure factor $g^{(1s)}(q,t)$ could generally be followed until $g^{(1s)}(q,t)/g^{(1s)}(q,0) < 3 \cdot 10^{-4}$, corresponding to a signal-to-noise ratio ca.\ 3000. Such precision is not generally found in experimental studies. In the subfield of microrheology, it is sometimes presumed\cite{dasgupta2002a} that $g^{(1s)}(q,t)$ is related to the mean-square displacement via
\begin{equation}
      g^{(1s)}(q,t) =  \exp(- q^{2} \langle(\Delta x(t))^{2} \rangle/2).
      \label{eq:g1scalc}
\end{equation}
This hypothesis was tested by plotting the directly-calculated (eq \ref{eq:g1s}) and inferred (eq \ref{eq:g1scalc}) values for $g^{(1s)}(q,t)$ against each other for various $q$ and $t$.

\section{Results}

Extended simulations were made on a probe that followed the Langevin equation and a probe that followed the Mori-Zwanzig equation with an exponential memory propagator. Test calculations were also made on two probes having exponential+power law memory propagators, to confirm that our results were not anomalies unique to the exponential. All three propagators yield qualitatively similar results. We treat in detail only the exponential memory propagator.

The Langevin simulation yielded all expected properties: $P(\Delta x,t)$ was a Gaussian at all times;  $K_{4}$ and $K_{6}$ were both very nearly zero. $C_{VV}(t)$ relaxed exponentially in t. At times sufficiently long that $C_{VV}(t)$ had relaxed into the noise in the simulation, $\langle (\Delta x(t))^{2} \rangle$ increased linearly with time, while $g^{(1s)}(q,t)$ was linear in $t$, linear in $q^{2}$, and was determined by $\langle (\Delta x(t))^{2} \rangle$ as seen in eq \ref{eq:g1scalc}.

\begin{figure}[ht!]
\subfigure{
\includegraphics[scale=0.7]{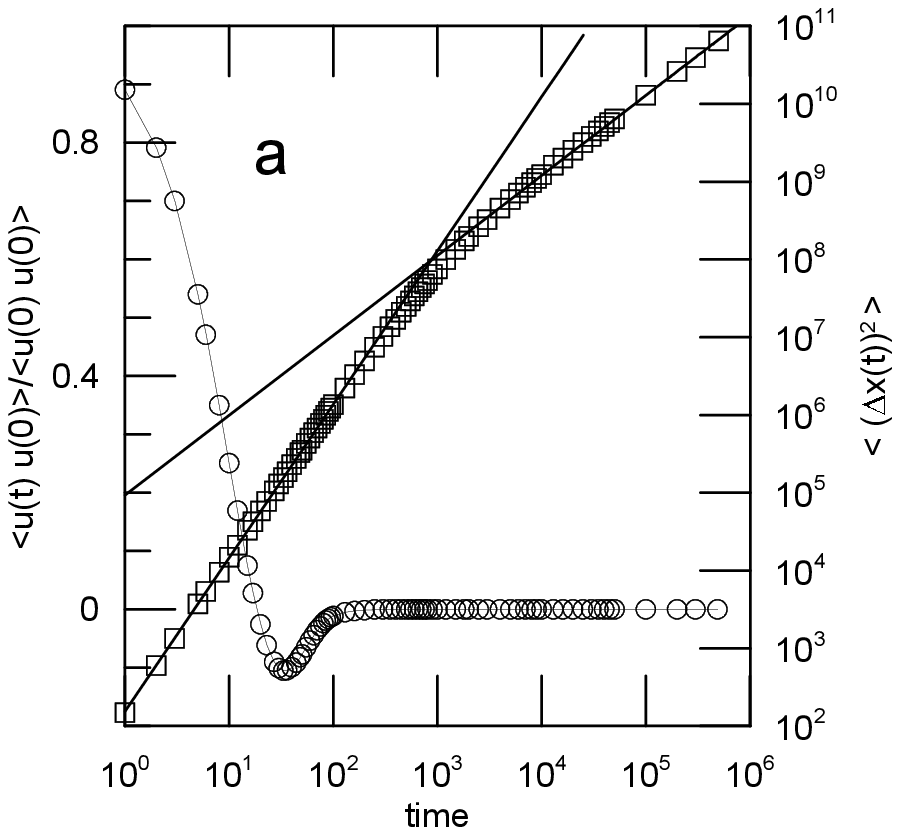}}
\subfigure{
\includegraphics[scale=0.7]{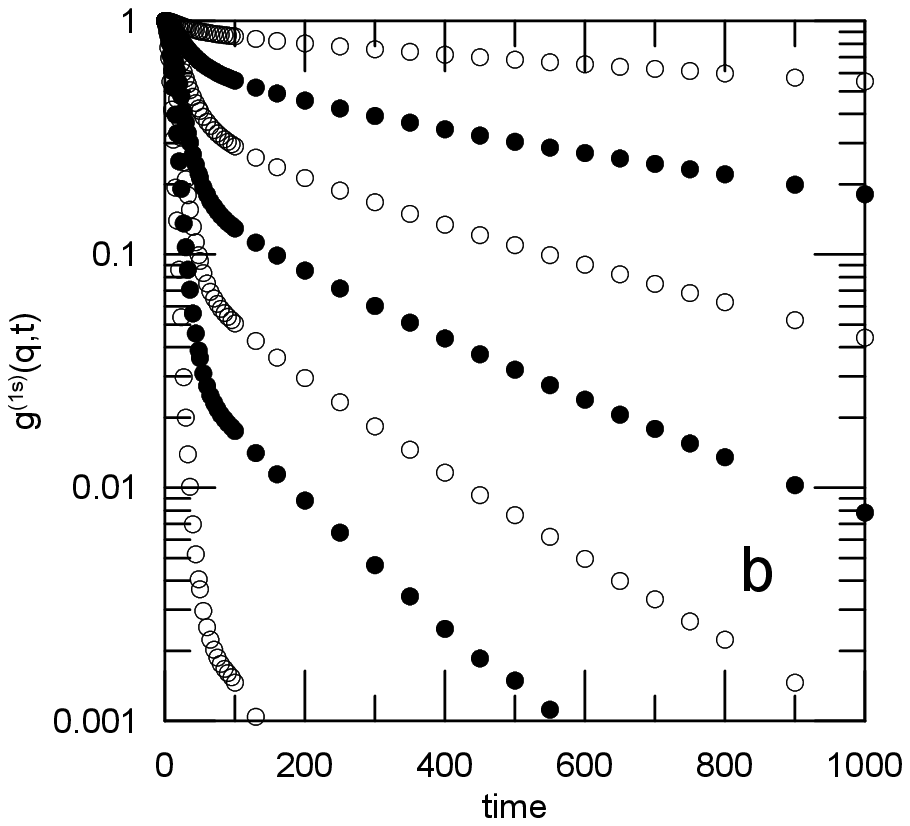}}
\subfigure{
\includegraphics[scale=0.7]{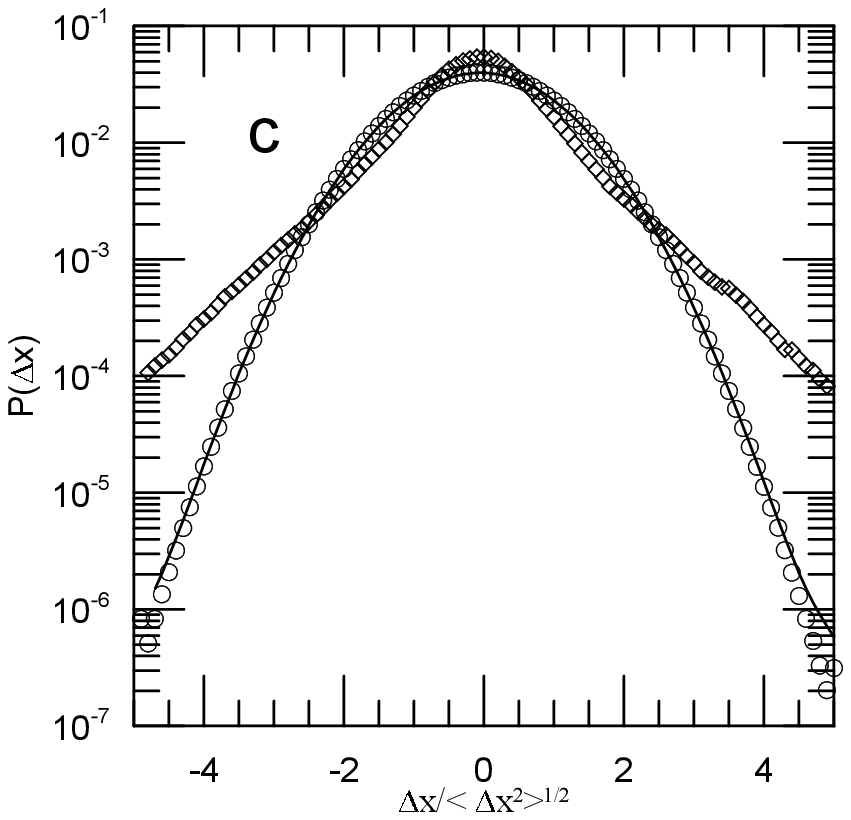}}
\subfigure{
\includegraphics[scale=0.7]{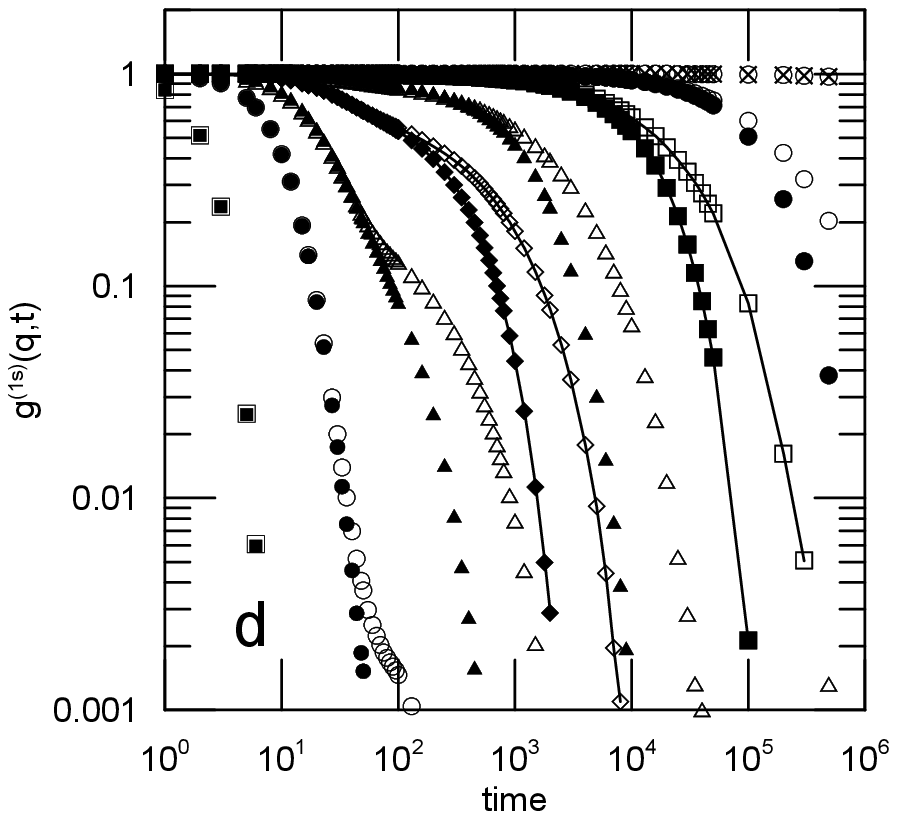}}
\caption{Dynamics of a Mori-Zwanzig walker having exponential memory.  (a) $\langle v(0) v(t) \rangle/\langle (v(0))^{2} \rangle$ ($\bigcirc$) and $\langle (\Delta x(t))^{2} \rangle$ ($\square$). (b)  $g^{(1s)}(q,t)$ for (from slowest to fastest decay) $q$ of 0.01, 0.02, 0.03, 0.04, 0.05, 0.06, and 0.08. (c) $P(\Delta x, t)$ at at 1 ($\bigcirc$) and 50,000 ($\lozenge$) timesteps. (d) Comparison of $g^{(1s)}(q,t)$ (open points) and $\exp(- q^{2} \langle (\Delta x(t))^{2} \rangle/2)$ (filled points), for $q$ (slowest to fastest decay) of 0.0001 ($\otimes$) , 0.001 ($\bigcirc$), 0.003 ($\square$), 0.01 ($\triangle$), 0.02 ($\lozenge$), 0.04 ($\triangle$),  0.08 ($\bigcirc$),  and 0.3 ($\square$).}
   \label{figureMExp}
\end{figure}

We now consider in detail the Mori-Zwanzig walker having an exponential memory propagator, eq \ref{eq:exppropagator}. Figure \ref{figureMExp} shows the important statistical properties of this probe.

The velocity-velocity correlation function and mean-square displacement appear as Fig \ref{figureMExp}a.
The time evolution of the mean-square displacement is undistinguished.  The two solid lines represent the  near-ballistic motion ($\langle (\Delta x(t))^{2} \rangle \sim t^{2}$) at short times and near-diffusive motion  ($\langle (\Delta x(t))^{2} \rangle \sim t^{1}$) at long times.  The evolution of $\langle u(0) u(t)\rangle$ shows a perhaps-unexpected feature, namely an oscillation resembling a damped ringing motion.  Damping is supplied by the hydrodynamic drag. The oscillation is driven by the memory kernel, which creates a drag proportional to the velocity (the sequential displacements) at earlier times.

A simple toy model demonstrates how the memory kernel leads to ringing in $C_{VV}(t)$.  Writing the velocity as $dx/dt$, and taking the memory kernel friction to act (the toy model) at a single earlier time $\theta$, the Mori equation assumes the form
\begin{equation}
     m\frac{d^{2}x(t)}{dt^{2}} = -M(\theta) \frac{dx(t- \theta)}{dt} - f_{o}\frac{dx(t)}{dt} + {\cal F}(t)+X(t)
      \label{eq:toymori}
\end{equation}
However, with an oscillatory solution $x \sim x_{o} \cos(\omega t)$, the velocity $dx(t-\theta)/dt \sim x_{o} \omega \sin(\omega (t - \theta))$ may be rewritten as a sum of terms $\sin(\omega t) \cos(\omega \theta) - \cos(\omega t) \sin(\omega \theta)$, allowing eq \ref{eq:toymori} with an appropriately chosen $\theta$ (one such that $\cos(\omega \theta)=0$) to be rewritten as
\begin{equation}
     m\frac{d^{2}x(t)}{dt^{2}} = - f_{o}\frac{dx(t)}{dt} + [M(\theta) \sin(\omega \theta)] x(t)+A(t)
      \label{eq:toymori2}
\end{equation}
with $A(t)$ being other time-dependent terms that are extraneous to the main result. For $M(\theta) \sin(\omega \theta) < 0$, as will be found with an appropriate $\omega \theta$, eq \ref{eq:toymori2} is very approximately the equation of a driven damped harmonic oscillator. The oscillations seen in $C_{VV}(t)$ (fig \ref{figureMExp}a) are thus explained.  In this simple case, the Mori-Zwanzig equation closely resembles the damped harmonic oscillator equation and has similar solutions. On making $f_{o}$ smaller (not shown), the oscillations in $C_{VV}(t)$ are found to become considerably more prominent.

Figure \ref{figureMExp}b shows the intermediate structure factor  $g^{(1s)}(q,t)$ as a function of time for various values of $q$. The relaxation of $g^{(1s)}(q,t)$ is profoundly non-exponential, with a drastic change in slope being apparent near $t =100$. The long-time relaxation of $g^{(1s)}(q,t)$ is not quite a simple exponential; it is seen to retain a slight curvature.

Figure \ref{figureMExp}c shows $P(\Delta x, t)$ at a series of times. There is an evolution in the qualitative shape of $P(\Delta x, t)$ between short ($t=1$) and long ($t=50000$) times. As seen in the figure, the short-time $P(\Delta x, 1)$ is a single Gaussian,  measurements (points) matching a Gaussian fit (solid line).  At large time $P(\Delta x, t)$ is not at all Gaussian.  The central feature in  $P(\Delta x, t)$, corresponding to  $\Delta x/\langle (\Delta x(t))^{2}\rangle^{1/2} < 1$ or so, is a central hump that could be approximated with a Gaussian. At larger $\Delta x$, $P(\Delta x, t)$ gains near-exponential wings, decreasing approximately as $\exp(- a | \Delta x|)$.

Fig \ref{figureMExp}d presents $g^{(1s)}(q,t)$ as a function of $t$ for various values of $q$. The purpose of the figure is to compare the measured $g^{(1s)}(q,t)$ with the Gaussian expectation $g^{(1s)} \sim \exp(- q^{2} \langle (\Delta x(t))^{2} \rangle/2)$.  At the largest $q$ reported, $g^{(1s)}(q,t)$ as measured agrees with the Gaussian expectation. However, at large $q$, $g^{(1s)}(q,t)$ decays into the noise at very small $t$.  At smaller $q$, the Gaussian expectation fails qualitatively.  At $q \leq 0.04$, the experimentally measured $g^{(1s)}(q,t)$ visibly becomes bimodal: At earlier times, $g^{(1s)}(q,t)$ agrees with the Gaussian expectation.  At later times, $\exp(- q^{2} \langle (\Delta x(t))^{2} \rangle/2)$ (filled points) falls rapidly with increasing $t$, while the measured $g^{(1s)}(q,t)$ (open points) decreases much more slowly.

\section{Discussion}

This paper reports a computer simulation of a probe particle whose motions are governed by an approximation to the Mori-Zwanzig equation.  The true Mori-Zwanzig equation is physically exact.  The approximation says the we are looking at a complex fluid that has a solvent component with rapidly relaxing fluctuations, and a solute component with a long-lived (here, exponential) relaxation.  The best test of the validity of our approximation is that it yields a calculated $P(\Delta x, t)$ that agrees with experiment, namely it has a central near-Gaussian hump and wings that relax as exponentials in $\Delta x$, exactly as observed experimentally by Wang, et al.\cite{wang2012a} and Guan, et al.\cite{guan2014a}.

We find that the Gaussian approximation eq \ref{eq:gaussianapprox} for $P(\Delta x, t)$ is incorrect for particles diffusing through a reasonable approximation to a complex fluid, as has also been seen experimentally.  $P(\Delta x, t)$ is not a Gaussian in $\Delta x$.  Correspondingly, $g^{(1s)}(q,t)$ is not a Gaussian in $q$. Over an extended range in $q$, $g^{(1s)}(q,t)$ is bimodal; $g^{(1s)}(q,t)$ at longer times is considerably larger than the Gaussian approximation prediction of eq \ref{eq:dynamicstructure2}.

I note several experimental techniques whose data interpretation sometimes relies on the Gaussian diffusion approximation, for which caution is therefore advisory.

Inelastic scattering methods, including quasielastic light scattering, quasielastic x-ray scattering, and inelastic neutron scattering, when applied to systems in which the scatterers are dilute, all measure $g^{(1s)}(q,t)$.  For each of these methods, the results above are all applicable.  Eq \ref{eq:dynamicstructure2} is invalid, at least in the system studied here.  If one used $g^{(1s)}(q,t)$ to infer the mean-square displacement, at long times and smaller $q$ the inferred mean-square displacement would be too small, and the inferred time-dependent microviscosity would be too large.

Pulsed-Field-Gradient nuclear magnetic resonance generally\cite{kuchel2012a} infers a self-diffusion coefficient via the Sjeskahl-Tanner equation\cite{stejskal1965a}, which in standard derivations\cite{kuchel2012a} inserts diffusion via the Fick's Second Law operator $D \nabla^{2}$, D being a constant.  The use of Fick's second law is equivalent to the Gaussian diffusion approximation.  Use of the Stejskal-Tanner equation and PFGNMR to infer self-diffusion coefficients of objects in complex fluids therefore requires careful attention. In particular, if the relaxation identified as corresponding to self-diffusion is not a simple exponential (cf.\ fig \ref{figureMExp}d), then the Gaussian diffusion approximation and hence the Stejskal-Tanner equation would not be applicable to the system.

Particle tracking techniques are sometimes only used to determine $\langle (\Delta x(t))^2\rangle$ rather than the full $P(\Delta x, t)$. If the mean-square displacement is interpreted directly as a time-dependent diffusion coefficient, the Gaussian approximation has been invoked implicitly, namely the relationship between $\langle (\Delta x(t))^2\rangle$ and $Dt$ is part and parcel of the Gaussian approximation.

Our findings, while assuredly not expected in parts of the complex fluids community, have extensive theoretical antecedents in other types of system. Haus and Kehr\cite{haus1987a} present a massive review of diffusion on regular and disordered lattices, including multiple sources demonstrating that $P(\Delta x, t)$ may have a decidedly non-Gaussian form, or that $\langle (\Delta x(t))^{2} \rangle$ may increase other than linearly in time. Bouchaud and Georges\cite{bouchaud1990a} present an extended discussion of \emph{anomalous diffusion} in which, e.g., the mean-square displacement shows subdiffusion or supradiffusion at long times.  Bouchaud and Georges emphasize that in order to see these effects some factor must intervene to cause the Central limit Theorem to become inapplicable. They note as effects causing anomalous diffusion the presence of long-range correlations, which lead to non-Markoffian steps in the random walk, anomalous dynamics leading to large fluctuations, and diffusion through quenched random media. Metzler and Klafter\cite{metzler2000a} consider diffusion through uniform media in systems described by fractional differential equations. They report particular conditions under which $P(\Delta x, t)$ that is far more sharply peaked than is a Gaussian. Srokowski and Kaminska\cite{srokowski2006a} treat a Markoff process that can exhibit subdiffusive or supradiffusive behavior. Luo, et al.\cite{luo1995a,luo1996a}, studied simple random walks in patterned and somewhat random static potentials, finding radically non-Gaussian forms for $P(\Delta x, t)$.

From a theoretical standpoint, it is thus not surprising that there are conditions under which particle diffusion in complex fluids is not Gaussian.

\end{document}